\documentclass[intlimits,twoside,a4paper]{article}

\usepackage{amsmath,amssymb}
\usepackage{graphicx}

\usepackage[T2A]{fontenc}
\usepackage[cp1251]{inputenc}

\usepackage{color}

\usepackage[eqsecnum]{cmpj2}

\issue{2016}{19}{2}{23002}
\doinumber{10.5488/CMP.19.23002}

\title[Ion clustering]%
{Ion clustering in aqueous salt solutions\\  near the liquid/vapor interface}
\author[J.D. Smith, S.W. Rick]{J.D. Smith, S.W. Rick}
\address{Department of Chemistry, University of New Orleans,
New Orleans, LA, 70148, USA}

\date{Received November 15, 2015}
\authorcopyright{J.D. Smith, S.W. Rick, 2016}

\begin{document}

\maketitle

\begin{abstract}

Molecular dynamics simulations of aqueous NaCl, KCl, NaI, and KI solutions are used to
study the effects of salts on the properties of the liquid/vapor interface.
The simulations use the models which include both charge transfer and
polarization effects. Pairing and the formation of larger ion clusters
occurs both in the bulk and surface region, with a decreased
tendency to form larger clusters near the interface.
An analysis of the roughness of the surface reveals that the
chloride salts, which have less tendency to be near the surface,
 have a roughness that is less than pure water, while
the iodide salts, which have a greater surface affinity, have a larger
roughness. This suggests that ions
away from the surface and ions
near the surface affect the interface in opposite ways.

\pacs 05.20.-y,  82.20.Wt,  83.10.Rs,   68.03.Hj
\end{abstract}

\section{Introduction}

Ions in aqueous solutions have different propensities to be
near the interface with the vapor phase \cite{PereraBerkowitz,StuartBerne,JungwirthTobias2001,Dang2002,Ghosal,PetersenSaykally,Jungwirth2006,Otten}.
The interest in the surface affinity  follows from the relevance to the environmental properties of the interface and from the
connection  to interfaces with electrodes and proteins \cite{JungwirthTobias2001,JungwirthWinter}.
The ions which have a greater affinity for the surface are larger, softer anions such as iodide and thiocyanate, while
smaller, harder ions (flouride or alkali cations) tend to avoid the interface.
The affinity for the surface is believed to
be driven not only by size but by polarizability \cite{PereraBerkowitz,StuartBerne,JungwirthTobias2001}.
The asymmetric solvation environment of the interface (relative to
the isotropic bulk) would favor the formation of a larger induced dipole on the ion, promoting the stability of
polarizable ions at the interface.
Recent studies have suggested a separate mechanism for surface selectivity, in which the ions stabilize the
surface entropically by enhancing fluctuations of the interfacial height \cite{Otten,Geissler2009,OuPatel2013}.
The ions with a surface affinity have a looser solvation shell which induces interfacial fluctuations larger than
either ions with tightly bound solvation shells or the pure liquid.

Entropic stabilization of the interface through enhanced fluctuations may also promote ion pairing at the
interface \cite{Garde2014}.
Ion pairing in bulk water is suggested in a number of experimental\cite{Fuoss,Aziz2006,MarcusHefter,Sebe} and
simulation \cite{BerkowitzKarimMcCammon,PettittRossky,Smith1994,PrattHummerGarcia,Bouazizi2006,SoperNaCl07,Fennell1,Fennell2,Timko,Ghosh,LuoRoux} studies.
Alkali halide salts show a small tendency to form pairs \cite{Hess2006,SoperNaCl07,LuoRoux,SoniatPairing}.
At a concentration of 1~M, the ions pair about 10 to 20$\%$ of the time, or only slightly above a random probability for pairing~\cite{SoniatPairing}.
Aggregation beyond pairs are observed as well \cite{Cerreta,Bian,Silva2000,ShermanCollings,ChenPappu,Hassan1,Hassan,SoniatPairing}.
The clusters tend to be small, most containing less than five ions, with a population that falls off exponentially with
size \cite{ChenPappu,Hassan,SoniatPairing}.
Aqueous KI appears to be different \cite{SoniatPairing}. Those ions pair with much higher probability and also form much larger clusters.
How the interface affects the clustering, beyond pairing, has not been examined.

In addition to polarization, charge transfer interactions have been shown to be important for the properties of ion
solvation \cite{Nadig1998,dalPeraro2005,Marenich2007,Varma2010,Zhao2010,Soniat2012,MgZnCT} and
aqueous interfaces \cite{Vacha2011,Vacha2012,Wick2012c,LeeAJ2012,Soniat2014,Roke14,Soniat15}.
Charge transfer between particles (water-water, ion-ion, and ion-water) can lead to charged solvation
shells around ions \cite{dalPeraro2005,Zhao2010,Soniat2012,Soniat2014,MgZnCT} and
charged interfaces \cite{Vacha2011,Vacha2012,Wick2012c,LeeAJ2012,Soniat2014,Roke14,Soniat15}.
The charge that is transferred between an ion and water is not exactly balanced by the charge
transferred with the counterion (typically, there is more charge transferred from the anion than is
transferred to the cation, \cite{Soniat2012} unless the cation is divalent \cite{MgZnCT}), resulting in
a net charge on the water molecules \cite{Sellner2013,Soniat2012,SoniatPairing}.
For alkali halides, each ion pair transfers about $-0.1 e$ of charge to the water.
Using the charge transfer, polarizable models, we will examine the amount of ion association near the
liquid/vapor interface for 1 molar solutions of NaCl, NaI, KCl, and KI.

\section{Methods}

{\bf The charge transfer model.} The simulations use a recent force field which includes both charge transfer and polarizability \cite{Lee2011,Soniat2012,SoniatPairing}.
Charge transfer is treated using the discrete charge transfer (DCT) method, in which charge is transferred between pairs based
on the distance. For water, a small amount ($- 0.02 e$) of charge is transferred from the hydrogen bond acceptor to the donor.
For anions, charge is transferred to water based on the distance between the ion center and the hydrogen on the water, and for
cations, it is based on the distance between the ion center and the oxygen atom.
Charge is transferred between unlike ions based on the distance between ion centers and no charge is transferred between ions of the
same charge. The charge transfer amounts are chosen to reproduce the results of quantum chemical calculations. Polarizability
is treated in the water molecules using the fluctuating charge formalism, in which charge can redistribute among atoms on the
same molecule in response to the electric field due to other atoms, and given a charge constraint determined by the
amount of charge transferred with other particles \cite{Lee2011}. For ions, polarizability is treated using a charge on a
spring, or Drude, model \cite{Soniat2014}.
The models have Lennard-Jones interactions between non-hydrogen atoms and Coulombic interactions between charge sites.

The water-water part of the potential was optimized to reproduce a number of properties of the liquid, including energy, density, pair correlation functions, dielectric
constant, and diffusion constant \cite{Lee2011}. As relates to this study, the model also accurately reproduces the surface tension \cite{Wick2012c}.
The ion-water potential was parameterized to reproduce single ion solvation free energies, coordination structure, average ion dipole and charge, in the liquid
phase, plus ion-water dimer properties \cite{Soniat2012,Soniat2014}.
Finally, the ion-ion interactions were adjusted to reproduce the osmotic pressure as a function of concentration \cite{SoniatPairing}.
The osmotic pressure is sensitive to the amount of ion pairing and provides a good property to optimize ion-ion interactions
against \cite{Hess2006,KalcherDzubiella,Fyta,LuoRoux,LuoRoux2}.

{\bf Simulation details.} The simulations use 2840 water molecules and 104 ions, creating a 1~M solution. An orthorhombic box of dimensions $44\times44\times176$~{\AA}, periodic in all directions. This creates a water layer about 44~{\AA}  thick, with a 132~{\AA} thick vapor layer.
Ewald sums were used for the electrostatic interactions, with a correction to mimic a system periodic in 2 dimensions \cite{YehBerkowitz}.
Additional details of the simulations are the same as previously published, \cite{Soniat2014,SoniatPairing}
in the $TVN$ ensemble, at 298~K, using a Nos\'{e}-Hoover thermostat and bonds constrained using SHAKE \cite{AllenCSL}.
The four different salt solutions were simulated for a total of 6 nanoseconds. The pure liquid, with 2840 water molecules, in a $44\times44\times176$~{\AA} box
was simulated for comparison to the salt solutions.

{\bf Data analysis.} The interface was characterized by fitting the water density as a function of the $z$-coordinant (the direction perpendicular to the interface) to a hyperbolic tangent
function,
\begin{equation}
\rho(z) = {\frac{1}{2}} (\rho_\text{L}  + \rho_\text{V} )
- {\frac{1}{2}} (\rho_\text{L}  + \rho_\text{V} ) \tanh[(z-z_0)/d],
\label{eq:tanh}
\end{equation}
where $\rho_\text{L}$ and $\rho_\text{V}$ are the liquid and vapor phase densities, respectively, $z_0$ is the position of the Gibbs dividing surface, and d is
the width. Interfacial widths are commonly reported as 10-90 thicknesses, $t_l$, the length over which the density changes from 90\% to 10\% of
$\rho_\text{V}$, which is equal to $2.197d$. Properties of the interface are calculated relative to $z_0$.
Another method to define the surface uses the instantaneous surface (INS), which accounts for the roughness of the surface \cite{Willard2010}.
This method takes a specific configuration of the system and generates a density using a Gaussian convolution of the heavy atoms. A position of the
interface on a three dimensional grid can be found from the location where the density drops to one half of its bulk value.
In this analysis, the ions as well as the water molecules are used to define the density \cite{Stern2013,Soniat2014},
and a Gaussian width of 2.4~{\AA} and a grid spacing of 1.0~{\AA} is used, as in previous studies \cite{Willard2010,Stern2013,Soniat2014}.
The INS method allows for characterization of the interface in terms of the fluctuations of the interfacial height \cite{Otten,Geissler2009,OuPatel2013}.
If the average height of the grid points corresponding to the INS is $\langle h \rangle$, the fluctuations in the height can be
found from $\langle \delta h^2 \rangle = \langle ( h - \langle h \rangle)^2 \rangle$. The shape of the interface can also be characterized by
the area of the INS \cite{Garde2014}. A useful quantity, $A_\text{excess}$, is the area of the instantaneous interface, $A$, divided by the area of the flat interface, or
$A_\text{excess}=A/(L_x L_y)$, where $L_x$ and $L_y$ are the box lengths in the $x$ and $y$ directions.
The ions are grouped into clusters based on pair distances \cite{SoniatPairing}.
Ions are taken to be paired if they are within the cut-off distance, $r_\text{cut}$, (3.5~{\AA}   for NaCl, 4.0~{\AA} for KCl and NaI, and 4.5~{\AA} for KI) and
are grouped into larger clusters if any ion is part of more than one pair \cite{ChenPappu,Hassan}.

\section{Results}

{\bf Properties of the interface.} The surface tension of the interface is found from the pressure tensor, as
\begin{equation}
\gamma = {\frac{L_z}{2}} \left[p_{zz}-{\frac{1}{2}} (p_{xx} + p_{yy} )\right],
\label{eq:surften}
\end{equation}
where $L_z$ is the box length in the $z$-direction (perpendicular to the interface) and $p_{\alpha \alpha}$ is the diagonal elements of the pressure tensor
in the $\alpha$ direction. Values of the surface tension are given in table~\ref{tab:properties}.
For comparison, the surface tension of pure water using TIP4P-FQ+DCT
model \cite{Wick2012c} is given.
The 10-90 thickness, $t_l$, is about the same for pure water, aqueous NaCl and KCl, and increases for KI and NaI.
From the instantaneous interface analysis, the fluctuations in the interfacial height, $\langle \delta h^2 \rangle^{1/2}$,
and the excess surface area, $A_\text{excess}$ can be determined.
The results show that for the NaCl and KCl solutions, $\langle \delta h^2 \rangle^{1/2}$ and $A_\text{excess}$ are less than they are for
pure water and for the NaI and KI solutions, $\langle \delta h^2 \rangle^{1/2}$ and $A_\text{excess}$  are about equal to the pure water values.
The 10-90 thickness, $t_l$, is about the same for pure water, aqueous NaCl and KCl, and increases for KI and NaI. This would be consistent with
more iodide than chloride ions at the surface.

\begin{table}[!h]
\caption{Surface tension, interfacial width, fluctuations in interface height, and excess surface area for water and the four 1~M solutions.\label{tab:properties}}
\vspace{2ex}
\begin{center}
\begin{tabular}{lcccc}
\hline
\hline
& $\gamma$ & $t_l$ & $\langle \delta h^2 \rangle^{1/2}$ & $A_\text{excess}$ \\
&  (dyn/cm)  & (\AA) & (\AA)  &  \\
\hline
pure water & 73$\pm$1 & 4.01$\pm$0.07 & 1.19$\pm$0.02 & 1.152$\pm$0.001   \\
NaCl & 89$\pm$2 & 4.05$\pm$0.05 & 1.16$\pm$0.01 & 1.148$\pm$0.001 \\
KCl & 87$\pm$1 & 4.01$\pm$0.06 & 1.17$\pm$0.02 & 1.149$\pm$0.001 \\
NaI & 76$\pm$2 & 4.33$\pm$0.11 & 1.21$\pm$0.02 & 1.152$\pm$0.001 \\
KI & 78$\pm$2 & 4.72$\pm$0.14 & 1.23$\pm$0.02 & 1.154$\pm$0.001 \\
\hline
\hline
\end{tabular}
\end{center}
\end{table}

\begin{figure}[!t]
\centering
\includegraphics[width=0.54\textwidth]{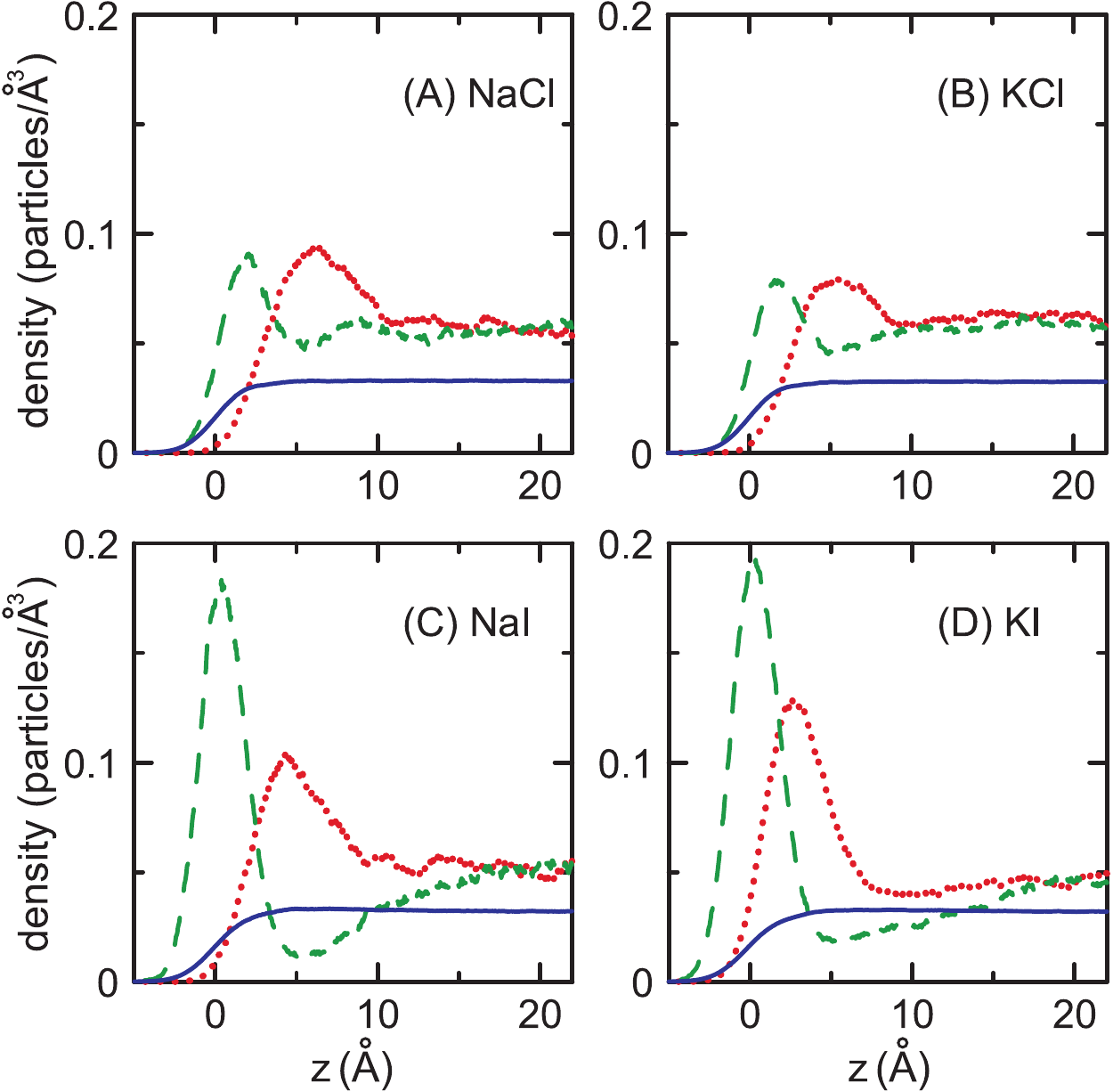}
\caption{(Color online) Density profiles as a function of the distance from the Gibbs dividing surface for the anion (green dashed line), cation (red dotted line) and water (blue
solid line). Values less than zero correspond the the vapor phase. The densities for the
ions have been multiplied by a factor of 100 to be on the same scale as the water densities.}
\label{fig:rho1}
\end{figure}

\begin{figure}[!b]
\centering
\includegraphics[width=0.54\textwidth]{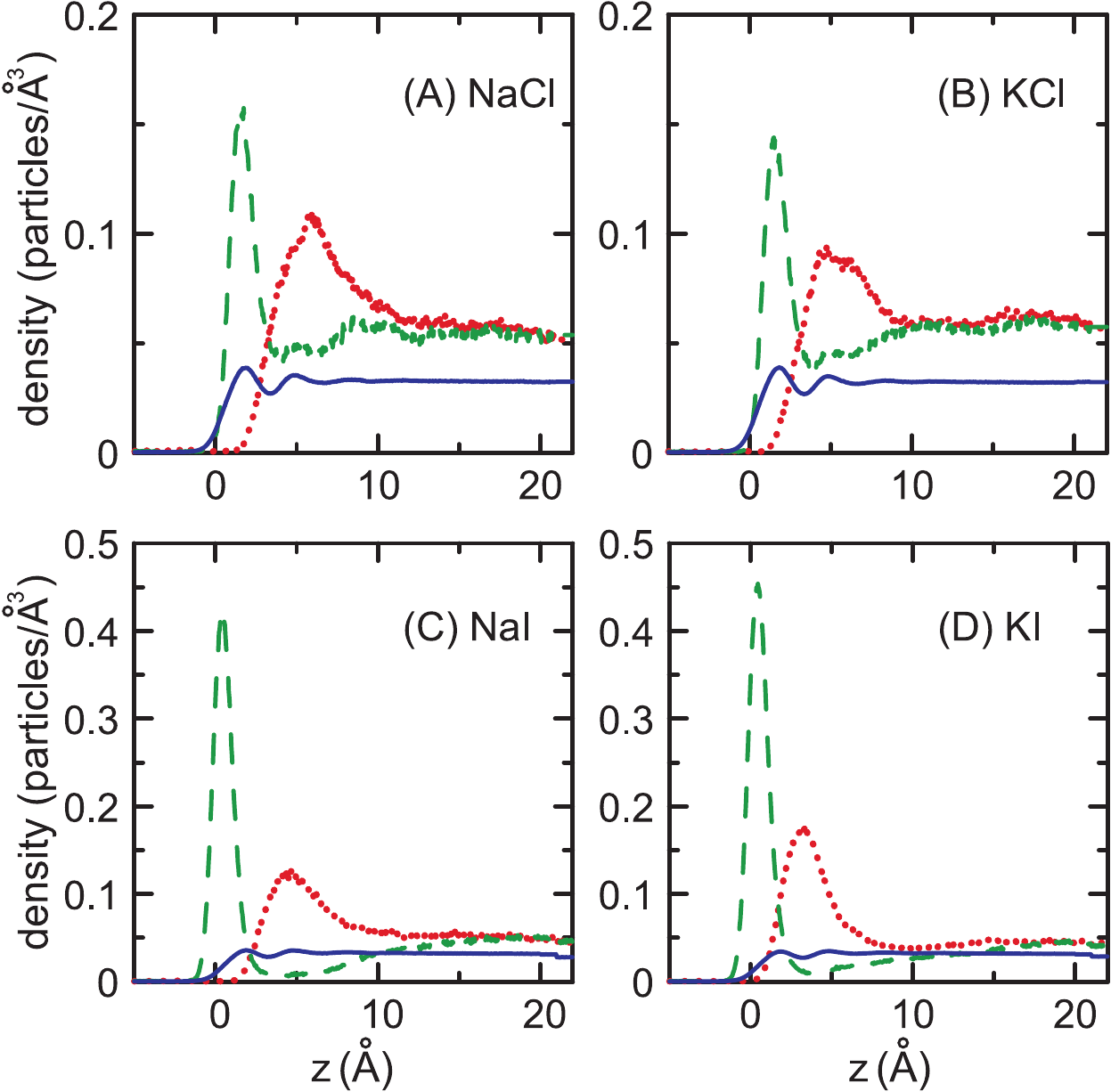}
\caption{(Color online) Density profiles as a function of the distance from the instantaneous surface for the anion (green dashed line), cation (red dotted line) and water (blue
solid line). Values less than zero correspond the the vapor phase. The densities for the
ions have been multiplied by a factor of 100 to be on the same scale as the water densities.}
\label{fig:rho1B}
\end{figure}

{\bf Ion densities.} The ion densities relative to the Gibbs dividing surface are shown in figure~\ref{fig:rho1}.
Iodide shows more surface affinity that chloride, and the cations occupy positions beneath away from the surface.
Figure~\ref{fig:rho1B} shows the density profiles relative to the instantaneous interface. When viewed this way,
the density profiles are sharper and the water density shows some structure \cite{Soniat2014,Bresme2012,Stern2013}.
The anions and cations are seen to be in distinct layers relative to the instantaneous interface.

\begin{figure}[!t]
\centering
\includegraphics[width=0.55\textwidth]{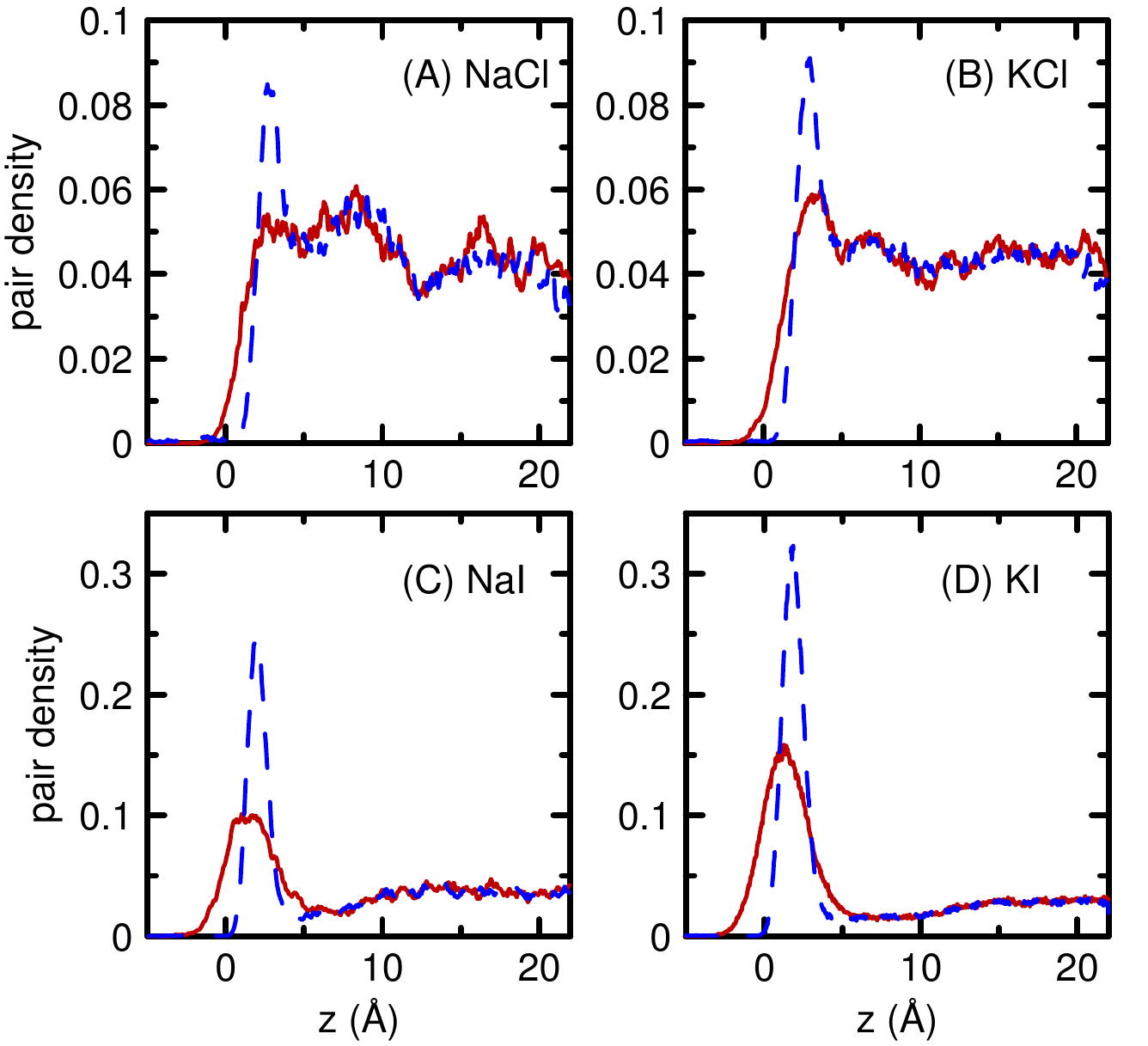}
\caption{(Color online) Pair probability as a function of the distance from the Gibbs dividing surface (red solid line) and instantaneous surface (blue dashed line).}
\label{fig:rhopair}
\end{figure}

{\bf Ion pairing.} The distribution of ion pairs is shown in figure~\ref{fig:rhopair}. Shown is the distribution of pairs, based on the center of mass of the pair, divided
by the total number of the pairs in the whole system. Enhanced pairing at the interface is apparent for all salts, as is especially evident from the distribution relative to
the instantaneous surface. Pairing with iodide is enhanced more than chloride, as might be expected since iodide has a greater surface affinity.
The increased amount of pairing may  arise since there are more anions at the surface or because
there are properties of the surface that promote pairing, as suggested by Venkateshwaran, {et al.} \cite{Garde2014}.

\begin{figure}[!b]
\centering
\includegraphics[width=0.55\textwidth]{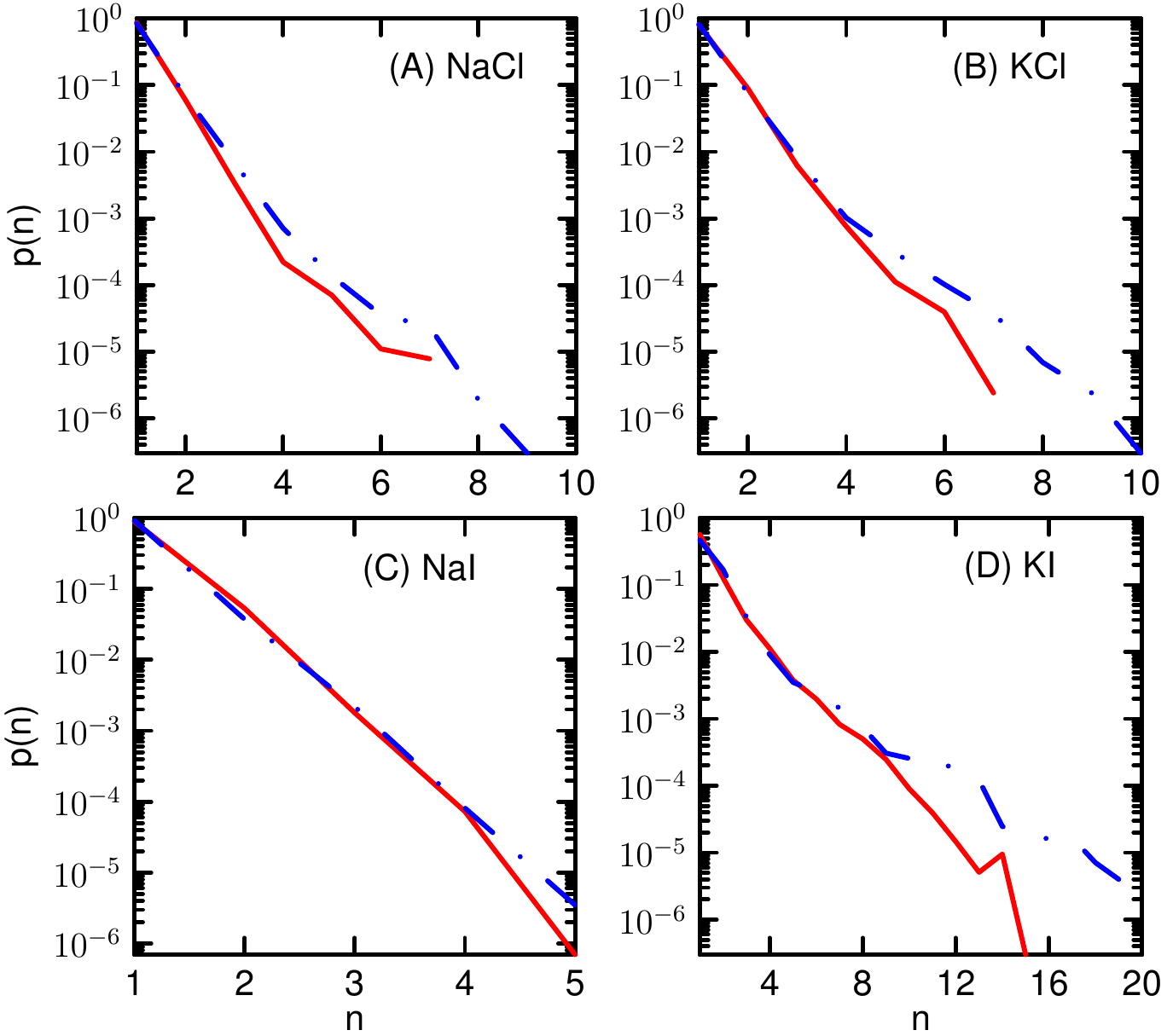}
\caption{(Color online) Distribution of clusters near the surface (red solid line) and the bulk (blue dashed line) regions.}
\label{fig:pcl}
\end{figure}

{\bf Ion clustering.}  To analyze how clustering changes at the surface, the system is split up into three regions, as suggested by
figures~\ref{fig:rho1B}  and \ref{fig:rhopair}. The $z$-coordinate of the center of the cluster relative to the instantaneous surface,
$z_\text{cluster}$, is used.
The first region ($z_\text{cluster}  <5$~\AA) corresponds to the surface, the second ($5~\text{\AA} <  z_\text{cluster}  < 12~\text{\AA}$), the
subsurface region, and the third ($z_\text{cluster} > 12$~\AA) corresponds to the bulk region. Within each region, cluster distributions, $p(n)$,
are determined. As done previously, \cite{SoniatPairing} the  distributions are normalized as
\begin{equation}
\sum_{n=1}^{N_\text{ion}} p(n) n = 1
\label{eq:pcluster}
\end{equation}
so that  $p(1) =1$ if all ions are present as single ions, $p(2) = 1/2$ if all are present as pairs, and $p(n)=1/n$ if all
ions are in a $n$-particle cluster. The distributions are shown in figure~\ref{fig:pcl}, comparing the surface to the bulk
region.
Larger clusters are seen with KI, consistent with the earlier study on bulk solutions \cite{SoniatPairing}.
There are differences between the bulk and the surface region, with less larger clusters at the
surface. On the scale of the plot, differences for small clusters are hard to discern, so the $p(n)$ values are
given in table~\ref{tab:pcl}. KCl and NaI show more, while NaCl shows slightly less and
KI significantly less pairing at the surface.

\begin{table}[!t]
\caption{Distributions of clusters of size one to three in the three different regions relative to
the interface. \label{tab:pcl}}
\begin{center}
\begin{tabular}{llccc}
\hline
\hline
& & p(1) & p(2) & p(3) \\
\hline
NaCl & surface & 0.87$\pm$0.02 & 0.060$\pm$0.009 & 0.003$\pm$0.001 \\
         & subsurface & 0.85$\pm$0.01 & 0.064$\pm$0.003 & 0.007$\pm$0.001 \\
         & bulk &            0.84$\pm$0.01 & 0.068$\pm$0.005  & 0.007$\pm$0.002 \\
 KCl & surface & 0.80$\pm$0.02 & 0.089$\pm$0.008 & 0.006$\pm$0.001 \\
        & subsurface & 0.81$\pm$0.01 & 0.078$\pm$0.004 & 0.008$\pm$0.001 \\
        & bulk        & 0.82$\pm$0.01 & 0.076$\pm$0.004 & 0.008$\pm$0.001 \\
 NaI & surface & 0.89$\pm$0.02 & 0.053$\pm$0.008 & 0.002$\pm$0.001 \\
       & subsurface & 0.93$\pm$0.01 & 0.031$\pm$0.002 & 0.002$\pm$0.001 \\
       & bulk &     0.92$\pm$0.02 & 0.037$\pm$0.007 & 0.002$\pm$0.001 \\
KI  & surface & 0.71$\pm$0.01 & 0.086$\pm$0.003 & 0.020$\pm$0.002 \\
      & subsurface & 0.65$\pm$0.02 & 0.089$\pm$0.005 & 0.023$\pm$0.003 \\
      & bulk & 0.49$\pm$0.02 & 0.163$\pm$0.002 & 0.032$\pm$0.002 \\
\hline
\hline
\end{tabular}
\end{center}
\end{table}

\begin{figure}[!b]
\centering
\includegraphics[width=0.4\textwidth]{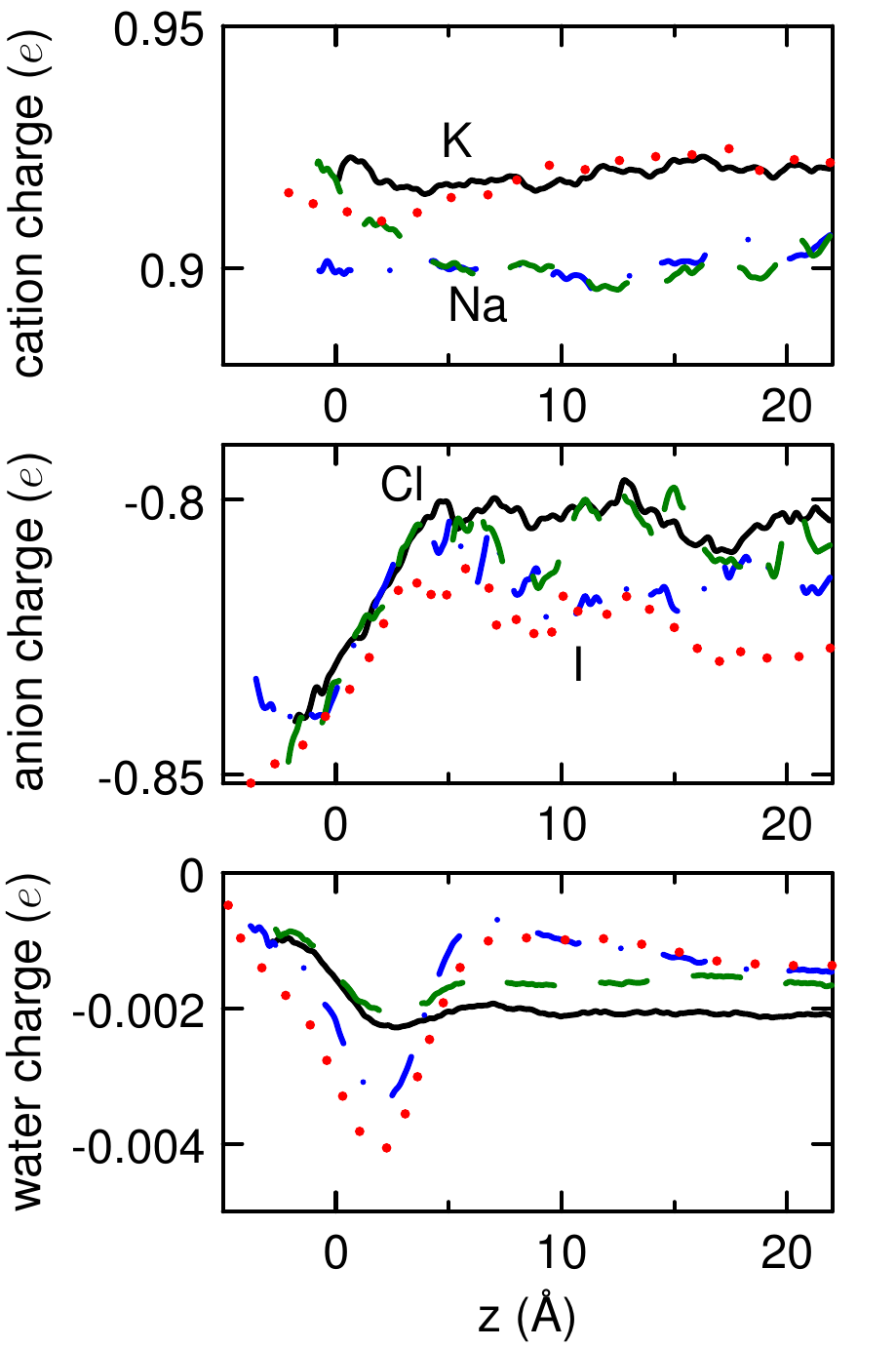}
\caption{(Color online) Charge of the cation (top), anion (middle) and water molecules (bottom) as a function of distance from the Gibbs dividing surface for NaCl (green dashed line),
KCl (black solid line),  NaI (blue dot-dashed line), and KI (red dotted line).}
\label{fig:ctvsz}
\end{figure}

{\bf Charge transfer effects.} The charge of the ions and the water molecules are shown in figure~\ref{fig:ctvsz}. The charge
transfer to the ions is less at the interface and both the anions and the cations have charges closer to their full charge.
The cations show a bigger change at the interface.
For the dilute ions, the DCT models show that the charge of the anions decreases at the interface, as the ions lose solvation shell water
molecules, while for cations, the charge is relatively unchanged \cite{Soniat2014}.
Pairing decreases the charge of the anion and increases the charge of the cation, as demonstrated with the DCT models \cite{SoniatPairing},
{\it ab initio} analysis of NaCl structures from classical simulations \cite{Sellner2013}, and {\it ab initio} molecular dynamics
studies of LiF in water \cite{Pluharova2013lett}.
Pairing leads to a less charge transfer, from the DCT perspective, because there is less charge transfer between an ion
pair than there is between an ion and water, particulariy for the anions. At the surface, both an increased pairing and a loss of
some of the solvation shell, for anions, leads to a less charge transfer.

In the bulk, the charge of the anions is around $-0.8 e$ and the cations is $0.9 e$, leading to an overall charge of the water
molecules. This arrises because more charge is transferred to the water from chloride or iodide than is transferred from
sodium or potassium. At a 1~M concentration, there is about 55 water molecules for every ion pair, so the charge imbalance
of $-0.1 e$ gets among 55 water molecules, giving each an average charge of about $-0.002 e$.
This charge on the water molecules, also seen in other simulation studies, \cite{Sellner2013,Soniat2012,SoniatPairing}
has implications on the dynamics of the water \cite{BerkowitzCT}.
At the surface, the water molecules can acquire a charge due to hydrogen bond
imbalances \cite{Vacha2011,Vacha2012,Wick2012c,LeeAJ2012,Soniat2014,Roke14,Soniat15}.
Molecules right at the surface tend to accept more hydrogen bonds than they donate to other molecules (often
termed ``dangling hydrogen bonds'') while molecules right beneath the surface donate more than they accept.
This gives rise to water molecules which are more negative near the surface (around $z=2$~{\AA}) and molecules
in the next layer (around $z=6$~{\AA}) that are more positive than bulk waters.

\section{Conclusion}

The simulation results find that the
surface tension depends most strongly on the anion, with a smaller surface tension for iodide. This is
in agreement with experiment, which finds that the surface tension increases as KI $\approx  \text{NaI} < \text{KCl} < \text{NaCl}$ \cite{Washburn}.
The increases are larger for the model, over 10~dyn/cm for NaCl for example, than they are experimentally (1.64~dyn/cm).
The surface tension for pure water varies by over 30~dyn/cm for different water models \cite{Wick2012c,VegadeMiguel,RouxDrude2}, so surface tension
is very sensitive to the details of the intermolecular interactions.
Simulations using non-polarizable models have found good agreement for the surface tension for aqueous NaCl \cite{ChenSmith,DAuria2009,Neyt2013},
while the results using polarizable models are not as good, showing a decrease in the surface tension \cite{Neyt2013}.
We are encouraged that the DCT models give the correct trend.

As in previous studies, with  DCT models at dilute concentrations \cite{Soniat2014}
and other polarizable potentials \cite{Jungwirth2006},
the iodide and chloride ions both show an affinity for the surface, with the
iodide ion showing more than chloride (figure~\ref{fig:rho1}).
In dilute solutions, sodium and potassium are repelled from the surface \cite{Soniat2014}.
In the 1~M solutions, the presence of the anion at the surface leads to a peak in the cation distribution beneath the surface, so that
there is an interfacial region that is neutral.
This peak is slightly enhanced in the KI solution.
The ion peaks become more narrow and higher and the water density has the structure, when they are plotted relative to the instantaneous interface (figure~\ref{fig:rho1B}), consistent with the previous studies \cite{Soniat2014,Bresme2012,Stern2013},
the ion peaks become more narrow and higher and the water density shows some structure.
From this analysis, it is apparent that the chloride ion surface peak
matches the water surface peak while the iodide peak is shifted towards the vapor phase. The sodium and potassium peaks are shifted more into the liquid phase than
they appear in figure~\ref{fig:rho1}.  In either analysis of the density, there is a region of depleted anion density beneath the surface, consistent with
other studies \cite{Warren2008JPhChC,DAuria2009}. For dilute anions, there is no such region \cite{Soniat2014}, indicating that this feature is induced
by the presence of the other ions.

The four solutions all show more pairing at the surface (figure~\ref{fig:rhopair}). The amount of pairing is a combination of
an increased density of anions and the tendency to pair.
That is, there might be more pairs because there are more ions or because
the surface promotes pairing, as has been proposed by
Venkateshwaran {et al.} \cite{Garde2014}.
The fraction for each ion to have no other ions in its solvation shell, making it a cluster of size one, is greater for NaCl and
KI, and slightly less for KCl and NaI (table~\ref{tab:pcl}). Relative to the amount of ions present, there is less pairing
for NaCl and KI.
For all four ion solutions, there are fewer larger clusters at the surface than in the bulk region (figure~\ref{fig:pcl}).
The tendency to form larger clusters is the greatest for KI, as reported earlier for the bulk \cite{SoniatPairing}.
For these solutions, there does not seem to be an enhanced tendency to form pairs, or larger clusters, at the surface,
possibly reflecting the differences between dilute and concentrated solutions.

The shape of the interface can be characterized by the fluctuations in the interfacial height, $\langle \delta h^2 \rangle^{1/2}$,
and the instantaneous surface area, as in previous studies \cite{Otten,Geissler2009,OuPatel2013,Garde2014}. Here, we
define the excess surface area, $A_\text{excess}$, as the instantaneous surface area divided by the surface area of the flat
interface.  Both these properties are measures of the deviations from a flat interface,
$\langle \delta h^2 \rangle^{1/2}$ would be zero and $A_\text{excess}$ would be one for a perfectly flat interface. They both increase
as the surface gets rougher. For NaCl and KCl, the surface is flatter than the pure solution (see table~\ref{tab:properties}).
A decrease in the surface area and in the surface height fluctuations is consistent with a large
increase in the surface tension of these two salts.
For iodide salts, the surface is rougher than the chloride salts, and $\langle \delta h^2 \rangle^{1/2}$ and $A_\text{excess}$ are
equal to the values for pure water.
Simulation studies on dilute ions suggest that a single ion \cite{Otten,Geissler2009,OuPatel2013}
or an ion pair~\cite{Garde2014} enhance surface fluctuations.
Combining those results with those of this study suggests that ions in the bulk decrease the fluctuations, and increase the surface tension, and
ions at the interface increase fluctuations, and decrease the surface tension.

\section{Acknowledgements}
This work was supported by the National Science Foundation under contract number CHE-0611679.


\ukrainianpart

\clearpage

\title{Іонне кластерування у водному розчині солі \\ поблизу границі розділу рідина/пара
}
\author{Ж.Д. Сміт, С.В. Рік}
\address{
Хімічний факультет, Університет Нью Орлеану, Нью Орлеан,
Луізіана 70148, США
}

\makeukrtitle

\begin{abstract}
\tolerance=3000%
Моделювання методом молекулярної динаміки водних розчинів NaCl, KCl, NaI, і KI використовується для дослідження ефектів солі на властивості границі розділу рідина/пара. Симуляції використовують моделі, які включають і зарядовий перенос, і поляризаційні ефекти. Парування і формування більших іонних кластерів має місце і в об'ємі, і в поверхневій області з тенденцією до зменшення формування великих кластерів біля границі розділу. Аналіз нерівності поверхні вказує, що хлоридні солі, які мають меншу тенденцію бути біля поверхні, мають нерівність, що є меншою, ніж у чистій воді, тоді як йодові солі, які мають більшу поверхневу спорідненість, мають більшу нерівність. Це передбачає, що іони здаля від поверхні та іони поблизу поверхні впливають на границю розділу протилежним чином.

\keywords границя розділу, перенос заряду, парування іонів, водні іони

\end{abstract}

\end{document}